# Moiré Potential Impedes Interlayer Exciton Diffusion in van der Waals Heterostructures

Junho Choi[1]*, Wei-Ting Hsu[1]*, Li-Syuan Lu[2], Liuyang Sun[1], Hui-Yu Cheng[2], Ming-Hao Lee[3], Jiamin Quan[1], Kha Tran[1], Chun-Yuan Wang[1,4], Matthew Staab[1], Kayleigh Jones[1], Takashi Taniguchi[5], Kenji Watanabe[5], Ming-Wen Chu[3], Shangjr Gwo[4], Suenne Kim[6], Chih-Kang Shih[1], Xiaoqin Li[1+], and Wen-Hao Chang[2,7+]

[1]Department of Physics, Complex Quantum Systems, and Texas Materials Institutes, The University of Texas at Austin, Austin, TX 78712, USA
[2]Department of Electrophysics, National Chiao Tung University, Hsinchu 30010, Taiwan
[3]Center for Condensed Matter Sciences, National Taiwan University, Taipei 10617, Taiwan
[4]Department of Physics, National Tsing-Hua University, Hsinchu 30013, Taiwan
[5]National Institute for Material Science, 1-1 Namiki, Tsukuba, Ibaraki 305-0044, Japan
[6]Department of Photonics and Nanoelectronics, Hanyang University, Ansan 15588, Republic of Korea
[7]Center for Emergent Functional Matter Science (CEFMS), National Chiao Tung University, Hsinchu 30010, Taiwan

* These authors contributed equally to this work
+Corresponding Authors:
elaineli@physics.utexas.edu
whchang@mail.nctu.edu.tw

**The properties of van der Waals (vdW) heterostructures are drastically altered by a tunable moiré superlattice arising from periodic variations of atomic alignment between the layers. Exciton diffusion represents an important channel of energy transport in semiconducting transition metal dichalcogenides (TMDs). While early studies performed on TMD heterobilayers have suggested that carriers and excitons exhibit long diffusion lengths, a rich variety of scenarios can exist. In a moiré crystal with a large supercell size and deep potential, interlayer excitons may be completely localized. As the moiré period reduces at a larger twist angle, excitons can tunnel between supercells and diffuse over a longer lifetime. The diffusion length should be the longest in commensurate heterostructures where the moiré superlattice is completely absent. In this study, we experimentally demonstrate that the moiré potential impedes interlayer exciton diffusion by comparing a number of WSe$_2$/MoSe$_2$ heterostructures prepared with chemical vapor deposition and mechanical stacking with accurately controlled twist angles. Our results provide critical guidance to developing "twistronic" devices that explore the moiré superlattice to engineer material properties.**

Van der Waals (vdW) materials provide exciting opportunities to create new heterostructures with vastly expanded choices of materials as the strict requirement of lattice matching between adjacent layers has been lifted[1]. Most intriguingly, a moiré superlattice emerges because of the periodic variations in atomic alignment between adjacent layers. The period of the moiré superlattice can be readily controlled by choosing materials with different lattice constants or adjusting the twist angle between the layers, leading to a new paradigm in engineering quantum materials[2, 3, 4]. For example, graphene bilayers can exhibit either superconducting or insulating phases driven by electron correlation controllable via the twist angle[5, 6, 7, 8]. Such a strategy has yet to be carefully explored in controlling semiconductor heterostructures built with transition metal dichalcogenides (TMD)[9, 10, 11, 12].

A type-II band alignment is typically found in a TMD hetero-bilayer (hBL), leading to rapid electron-transfer to one layer and hole-transfer to the other layer, and the formation of interlayer excitons ($IX$s)[13, 14, 15]. These $IX$ resonances are the lowest energy optical excitations and shifted from the intra-layer excitons mainly by the band alignments[16, 17, 18, 19, 20, 21, 22, 23, 24, 25]. The diffusion and mobility of various carriers including electrons, holes, and $IX$s are important properties because they determine the charge and energy transport processes in hBLs, which, in turn, are critical for building transistors and photovoltaic devices. Several experiments performed on $WS_2/WSe_2$ and $MoS_2/WSe_2$ hBLs have found long free carrier diffusion length[26] and demonstrated electric-field control of the $IX$ transport at room-temperature[23]. Other recent experiments, on the other hand, suggest that $IX$s may be localized by the moiré potential on the order of 100-200 meV[24, 25]. In fact, these seemingly conflicting results suggest rich opportunities to control the optical properties of TMD heterostructures.

In this paper, we investigate the influence of the moiré potential on $IX$ diffusion in $WSe_2/MoSe_2$ hBLs using spatially- and time-resolved photoluminescence (PL) measurements. We carefully compare experiments performed on two types of samples either grown by chemical vapor deposition (CVD) or created by the mechanical exfoliation and transfer (MET) with a well-controlled twisted angle. $IX$ diffusion length in the rotationally aligned CVD grown hBL exceeds the size of the heterostructure and is estimated to be a few microns. In contrast, no $IX$ diffusion beyond the excitation laser spot is observed in the stacked hBL with ~ 1° twist angle, suggesting that $IX$s may be localized by the moiré potential. In another MET sample with ~ 3.5° twist angle,

*IX*s diffusion is observed but with a short diffusion length ~ 1 μm despite nearly ~ 100 fold longer lifetime than that in the CVD grown sample. These experiments suggest that *IX* diffusion is highly controllable via engineering the moiré superlattice and provide valuable guidance to extensive effort in searching for *IX* Bose-Einstein condensate[27, 28] or localized quantum emitters in the TMD heterostructures[3, 4, 24, 25, 29, 30].

**Experimental Results**

We briefly describe the samples investigated here. We focus on WSe$_2$/MoSe$_2$ hBLs with *R*-stacking style or near this stacking style where a deep moiré potential is expected[31, 32]. The stacking style is confirmed by transmission electron microscope images or second harmonic generation (SHG) (Supplementary information SI). The optical microscope images and structural illustrations for the CVD (sample *A*) and MET (sample *B* and *C*) methods are shown in Fig. 1a-b and 1d-e, respectively. The lattice constants of MoSe$_2$ and WSe$_2$ are very close to each other. In sample *A* grown by the CVD method, the lattice in each layer is slightly distorted to form a commensurate and rotationally aligned (i.e., 0° twist angle) hBL as illustrated in Fig. 1c. The hBL regions are limited to the bright-inner triangle with a typical lateral size of ~ 10 μm (Fig. 1a). In the MET samples, a twist angle $\theta = 1.1 \pm 0.3°$ (sample *B*) and $3.5 \pm 0.3°$ (sample *C*) from the *R*-stacking leads to the formation of a moiré superlattice (Fig. 1f). The twist angle is determined from a high-resolution optical image, SHG (see supplementary information (SI)) and further confirmed by *IX* lifetime measurements presented later. The presence of the moiré potential and its periodicity plays a major role in determining exciton diffusion as illustrated in Fig. 1c, 1f and demonstrated in the rest of the paper. Both intra-layer excitons and *IX*s are observed (PL spectra included in SI) while we focus on the properties of the *IX*s in this paper.

*IX* diffusion is investigated using spatially- and spectrally resolved PL in all three samples as shown in Fig. 2a-f. We examine the spatial PL line profiles at a few selected energies at the center, lower, and higher energy side of the broad *IX* spectral feature. No pronounced energy dependence is observed in any of these samples. However, the spatially *IX* diffusion is remarkably different between these three samples. The diffusion length in the CVD sample *A* is the longest and limited by the physical size of the heterostructure region. In contrast, no *IX* diffusion beyond the excitation laser spot size is observed in the MET sample *B* with $\theta = $ ~ 1°. The full-width at half-maximum

(FWHM) of a Gaussian function fitted to the excitation laser spot yields ~ 1 μm. The size of the *IX* PL spot appears slightly larger than the excitation laser spot size in Fig. 2e, which is caused by the imaging optics (e.g., microscope objective and curved mirrors inside the spectrograph) at different wavelengths (i.e., *IX*s near 900 nm and the excitation laser at 660 nm). Interestingly, the *IX* diffusion is found to depend on the moiré period as well as the *IX* lifetimes. When the moiré period is reduced from 20 nm in sample *B* to 5.7 nm in another MET sample *C* with $\theta = \sim 3.5°$, *IX* diffusion beyond the excitation spot size becomes observable as shown in Fig 2c and 2f.

*IX* diffusion can be quantitatively analyzed using the following equation[33, 34, 35]

$$\frac{\partial n(r,t)}{\partial t} = -\nabla \cdot \left(-D_{IX}\nabla n(r,t) - n(r,t)\mu u_0 \nabla n(r,t)\right) - \frac{n(r,t)}{\tau_{IX}} + G(r,t) \tag{1}$$

where $n(r,t)$ is the *IX* density, $D_{IX}$ is the diffusion coefficient, $\mu$ is the exciton mobility, $u_0$ is the *IX* interaction energy, $\tau_{IX}$ is the *IX* lifetime, and $G(r,t)$ is the exciton generation term, respectively (more details in SI). To avoid the complications arising from exciton-exciton annihilation and repulsive interaction at high excitation density[36, 37, 38, 39], all experiments presented here are performed at a low excitation power of 1 μW, corresponding to an estimated *IX* density of ~ $10^9$ /cm². We present the power-dependent studies for all three samples in SI. Making the assumption that the interaction effect can be neglected at low density, the steady-state *IX* density follows the analytical solution[36]:

$$n(r) \propto \int_{-\infty}^{\infty} K_0\left(\frac{r}{L_{IX}}\right) e^{-(r-r')^2/w^2} dr', \tag{2}$$

This solution is a convolution between the Gaussian profile of incident laser and the modified Bessel function of the second kind $K_0$ with the diffusion length $L_{IX} = \sqrt{D_{IX}\tau_{IX}}$, where $\tau_{IX}$ and $D_{IX}$ are the *IX* lifetime and diffusion coefficient, respectively. A series of solutions are generated by varying $L_{IX}$ as shown in Fig. 3a. Comparing the $1/e$ width of the simulated curves and the experimental data, we extracted the range of diffusion length for sample *A* to be ~ 3-16 μm and for sample *C* to be ~ 1 μm. The uncertainty in sample *A* partially arises from the truncated data due to the limited size of CVD grown flake. We used several different models to analyze $L_{IX}$ and the range of values is shown in Fig. 3b (see SI for more details).

We further analyze the diffusion coefficients of *IX*s by measuring *IX* lifetimes using time-resolved PL (TRPL) measurements. The data in Fig. 3c is fitted with either a bi-exponential or single-exponential decay function and extracted lifetimes are displayed in Fig. 3d. In sample *A*, a bi-exponential fit leads to a fast and slow component of $\tau_1^A = 1.2\ ns$ and $\tau_2^A = 5.5\ ns$ and a weighted average lifetime of 1.7 ns. This *IX* lifetime is similar to $\tau^B = 0.8\ ns$ for sample *B* but drastically different from $\tau^C = \sim 100\ ns$ for sample *C*. The lifetimes measured from both types of hBLs are consistent with the previous experiments[16, 25, 40, 41]. The significantly longer lifetime found in sample *C* is related to the momentum space shift between the conduction and valence bands in twisted bilayers, leading to an indirect transition in both real- and momentum space[42]. This angle-dependent lifetime is confirmed in many of our hBLs prepared by the MET method. Using $D_{IX} = L_{IX}^2/\tau_{IX}$, an effective diffusion coefficient is estimated to be ~ 50 cm²/s for sample *A*, larger than the typical value reported for incommensurate hBLs[21, 26, 43]. The extracted value for sample *C* is ~ 0.1 cm²/s, consistent with previous studies (see SI for details).

The absence of the moiré pattern in sample *A* is directly proven by a high-resolution STEM image as shown in Fig. 4a. The brightest points identify the positions of Mo atoms, the second brightest points correspond to W atoms, and the dimmest ones are from the Se atoms. The commensurate atomic registry is proven by the uniform pattern observed in the STEM image. The schematic in Fig. 4b illustrates the *3R*-like stacking of this atomic registry, in which the Se atoms are directly positioned on top of the Mo atoms. A MET hBL cannot be easily transferred and identified on a standard TEM grid. Thus, we perform atomic force microscopy (AFM) measurements to examine the atomic alignment in a mechanically stacked hBL. We follow the same procedure to prepare a sample without the top hBN to expose the WSe$_2$/MoSe$_2$ hBL, which enables AFM imaging (see Method for details). Despite some contaminants on the sample, we observe a disordered hexagonal pattern with a period of ~ 20 nm, suggesting a small twist angle of ~ 1° in this particular sample (Fig. 4c).

**Discussion**

Recognizing that the quality of hBLs can vary, we repeated these measurements on different samples and on different locations of the same sample (more data in SI). All experiments yielded consistent observations. The longest diffusion length is always observed in the CVD grown commensurate WSe$_2$/MoSe$_2$ hBLs. A finite diffusion length of ~ 1 μm is observed in a stacked hBL with a relatively large twist angle while no diffusion beyond the laser spot is observed in stacked hBLs with a small twist angle ($\theta = $ ~ 1°).

The strikingly different *IX* diffusion in CVD and MET hBLs is most naturally explained by taking into account the absence/presence of the moiré potential. While there are other differences between CVD grown and MET hBLs, e.g., substrates and defect density, these differences cannot account for the different *IX* diffusions. The CVD samples are not grown on an atomically smooth substrate and likely exhibit higher defect density. Both factors should lead to a shorter diffusion length than the MET samples[39]. Furthermore, the significant longer *IX* lifetime in the stacked hBL with $\theta = $ ~ 3.5° (sample *C*) should lead to a longer diffusion length than the CVD sample *A* if their diffusion coefficients were similar.

The difference between two MET hBLs arises from the expected changes in moiré periods and measured exciton lifetimes. For a moiré crystal with a small periodicity, excitons may tunnel between different supercells, and diffusion is observable over a long *IX* lifetime. For a moiré crystal with a relatively large periodicity, the tunneling between different supercells is exponentially reduced. *IXs* become localized within a moiré supercell due to a reduced lifetime and decreased velocity. Our conclusion is qualitatively consistent with previous experiments[21, 23, 26, 43] that reported finite *IX* or carrier diffusion in hBLs with a small moiré period in the sense exciton diffusion beyond the diffraction limit can be observed. Although some quantitative variations in diffusion lengths and diffusion coefficients exist in the literature, they mostly arise from the different diffusion models and analysis (SI) and, to a less extent, the preparation procedure of MET bilayers. To establish a more detailed correlation between the moiré supercell size and exciton diffusion, future experimental techniques with higher spatial and/or temporal resolutions are required.

**Conclusion**

Moiré potential has been demonstrated to have a profound influence on the electronic and optical properties of vdW heterostructures. However, its role in exciton diffusion has been largely ignored in previous studies on hBLs consisting of TMD monolayers with rather different lattice constants or with larger twist angles[21, 23, 43]. Our experiments present a complementary view of *IX* diffusion from previous studies and highlight the influence of the moiré potential on exciton diffusion in hBLs. These studies provide critical guidance for intensive efforts on searching for *IX* BEC or quantum emitters in TMD heterostructures. *IX* BEC only occurs when the exciton density exceeds a threshold value. In this case, mobile excitons are preferred and are more likely found in either CVD grown samples or MET samples with small moiré periods. On the other hand, localized excitons leading to a regular array of quantum emitters are more likely hosted in stacked hBLs with a small twist angle and large moiré period.

## Methods

### Optical measurements:

For the CVD grown sample *A*, PL measurements were taken at T = 4 K inside a cryogen-free low-vibration cryostat equipped with a three-axis piezo-positioner, and an objective lens (N.A. = 0.82) mounted inside the cryostat. The excitation sources at 632 nm (750 nm) were provided by a He-Ne (continuous-wave, CW Ti: Sapphire) laser. The PL signal was then sent to a 0.75 m monochromator and detected by a nitrogen-cooled charge-coupled device (CCD) camera. For TRPL measurements, a 635-nm pulsed laser diode (~90 ps width, 40 MHz repetition rate) was used as the excitation source, and the PL was detected by a fast avalanche diode. The PL decays were recorded using the time-correlated single-photon counting (TCSPC) technique with a temporal resolution of ~100 ps. Slightly different experimental conditions for measurements performed on sample *B* and *C* can be found in the SI.

**CVD growth process and STEM characterizations:** Single-crystal $WSe_2/MoSe_2$ hBLs were grown on sapphire substrates by CVD following a previously developed one-step growth process[44, 45]. High-purity $MoO_2$ (99%, Aldrich), $WO_3$, and Se powders (99.5%, Alfa) were used as the initial reactants. The hBLs were grown at 880°C in $H_2$/Ar flowing gas at low pressure (5–40 Torr). During the growth, the flow rates for $H_2$/Ar gas were set to 6/60 sccm. For STEM measurements, the samples were capped with poly(methylmethacrylate) (PMMA) (950K A4) by spin-coating and then baked at 100°C for 60 min. The $WSe_2/MoSe_2$ hBLs were then transferred onto a transmission electron microscopy (TEM) grid by PMMA-based wet transfer procedures. ADF-STEM images were obtained using a spherical aberration-corrected transmission electron microscope (JEOL-2100F). The detailed growth and characterization processes can be found in ref. [32].

**AFM measurements:** We obtained the phase image using the tapping mode of AFM (NX10, Park Systems) with a conductive cantilever (PPP-EFM) at ~ 28°C and ~ 68% relative humidity. To obtain the image, the tip was brought to the sample as close as possible, i.e., if the setpoint was lowered further, a large oscillating noise appeared due to the tip-sample interaction. The hexagonal pattern was only observable in selected areas of the sample due to surface contaminants. The inverse FFT of an AFM phase image was obtained using XEI (Park Systems) software.

**Mechanically stacked bilayer preparation:** The $MoSe_2$, $WSe_2$ monolayers (2D semiconductors), and the top hBN (hexagonal boron nitride) capping layer were mechanically exfoliated onto a polydimethylsiloxane (PDMS) sheet. The bottom hBN layer was directly exfoliated onto a $SiO_2$/Si substrate. The crystal orientations of $MoSe_2$ and $WSe_2$ monolayers were identified by polarization-resolved SHG. Using a hemisphere PDMS dot with polycarbonate (PC) film based dry-transfer technique[46, 47], the top hBN layer was picked up and used to stack $MoSe_2$ and $WSe_2$ bilayer by picking up each layer sequentially. The rotational alignment of the crystal orientations of the monolayers was guided by high-resolution optical microscope images. The stacked hBN/$MoSe_2$/$WSe_2$ layers were transferred on the bottom hBN layer, and then the heterostructure was dipped in an acetone bath to clean off PC residue. Thermal annealing in a high vacuum (~$10^{-7}$ mbar) at 200 °C for 4 hours was performed to improve the interfaces. Polarization-resolved SHG measurements on the heterostructure region were performed to confirm the twist angle and distinguish between *R*- and *H*-stacking orders [48].

**Data availability**
The data used to produce the figures in this paper are available from the corresponding authors upon reasonable request.

**Acknowledgments:** The spectroscopic experiments performed by J.C. at UT-Austin and X. Li were supported by the Department of Energy, Basic Energy Science program via grant DE-SC0019398. The partial support for K.T. was provided by the NSF MRSEC program DMR-1720595, which also facilitated the collaboration between the group of C.S. and X.L. L.S., C. S. was supported by the Welch Foundation via grant F-1662 and F-1672. M.S. and K. J. were supported by NSF EFMA-1542747. C. S. acknowledges support from the U.S. Air Force via grant FA2386-18-1-4097. W.C. acknowledges the support from the Ministry of Science and Technology (MOST) of Taiwan (105-2119-M-009-014-MY3 and 107-2112-M-009-024-MY3). W.H. acknowledges the support from the Ministry of Science and Technology of Taiwan (MOST-107-2917-I-564-010). M.L. and M.C. acknowledge the support from MOST. S.G. and C.W. acknowledge the support from the Ministry of Science and Technology (MOST) in Taiwan (MOST 105-2112-M-007-011-MY3). J.Q. acknowledges the support from the China Scholarship Council (Grant No.201706050068). K.W. and T.T. acknowledge support from the Elemental Strategy Initiative conducted by the MEXT, Japan and the CREST (JPMJCR15F3), JST. S.K. was financially supported by the National Research Foundation (NRF) of Korea grant funded by the Korea Government (2017R1D1B04036381). The collaboration between National Tsing-Hua University and The University of Texas at Austin is facilitated by the Global Networking Talent (NT 3.0) Program, Ministry of Education in Taiwan.

**Author contributions:** J.C. and W.H. led the optical experiments. L.S., H.C., and C.W. assisted the experiments. J.C. and L.L. led the sample preparations. K.T., M.S., K.J., assisted in making the samples. T.T. and K.W. provided hBN samples. M.L., W.H., and L.L. performed STEM measurements and analysis. J.Q. and S.K. performed AFM measurements and analysis. J.C., W.H., X.L., and W.C. wrote the manuscript. X.L., M.C., S.G., S.K., C.S., and W.C. supervised the project. All authors discussed the results and commented on the manuscript at all stages.

**Competing interests:** The authors declare no competing interests.

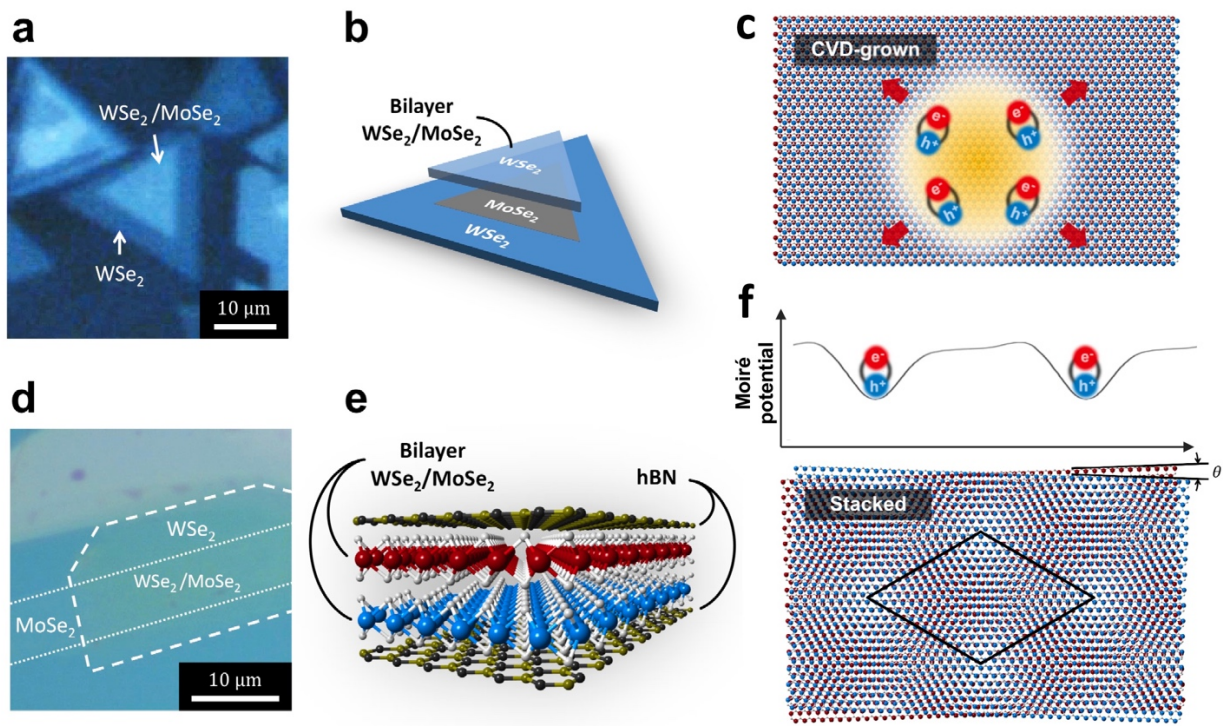

**Fig. 1. Two types of WSe$_2$/MoSe$_2$ heterostructures and the absence/presence of moiré superlattice.** Optical images of (a) CVD-grown sample *A* and (d) mechanically-stacked WSe$_2$/MoSe$_2$ hBL sample *B*. (b) Illustrated CVD-grown hBL, which consists of a monolayer WSe$_2$ covering on top of a monolayer MoSe$_2$-WSe$_2$ lateral heterostructure. (e) Illustration of the hBN encapsulated WSe$_2$/MoSe$_2$ hBL. (c) Schematics of exciton diffusion in a CVD grown WSe$_2$/MoSe$_2$ bilayer with a commensurate structure and (f) impeded exciton diffusion by the moiré potential in a MET hBL. The moiré supercell size decreases with an increasing twist angle $\theta$ between the two layers.

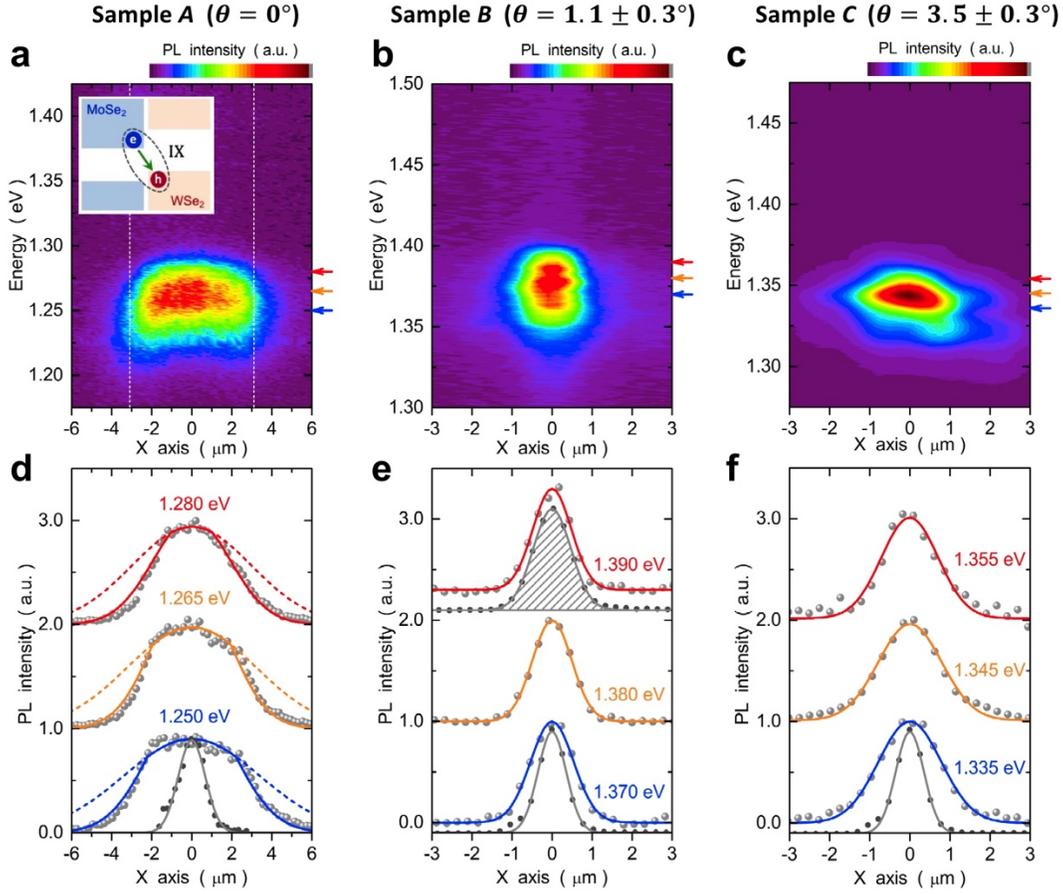

**Fig. 2. Spatially-resolved PL images.** 2D PL images and spectra of the *IXs* with the spatial (energy) coordinate shown in the horizontal (vertical) axis for (a) CVD-grown sample *A*, (b) MET sample *B* with $\theta = 1.1 \pm 0.3°$, and (c) MET sample *C* with $\theta = 3.5 \pm 0.3°$. A much longer *IX* diffusion length in sample *A* is clearly observed. The corresponding PL line profiles taken from (d) sample *A*, (e) sample *B* and (f) sample *C*, respectively. Note the PL profiles from sample *A* are truncated by the boundaries of the heterostructure region as indicated by the white dashed lines in (a). In addition, line profiles of excitation laser spots (bottom gray lines) are displayed. The 900-nm laser profile (top gray shaded) is displayed in (e) to illustrate that the PL image is limited by excitation spot size and imaging optics in sample *B*. Measured data points are fitted by a Gaussian function.

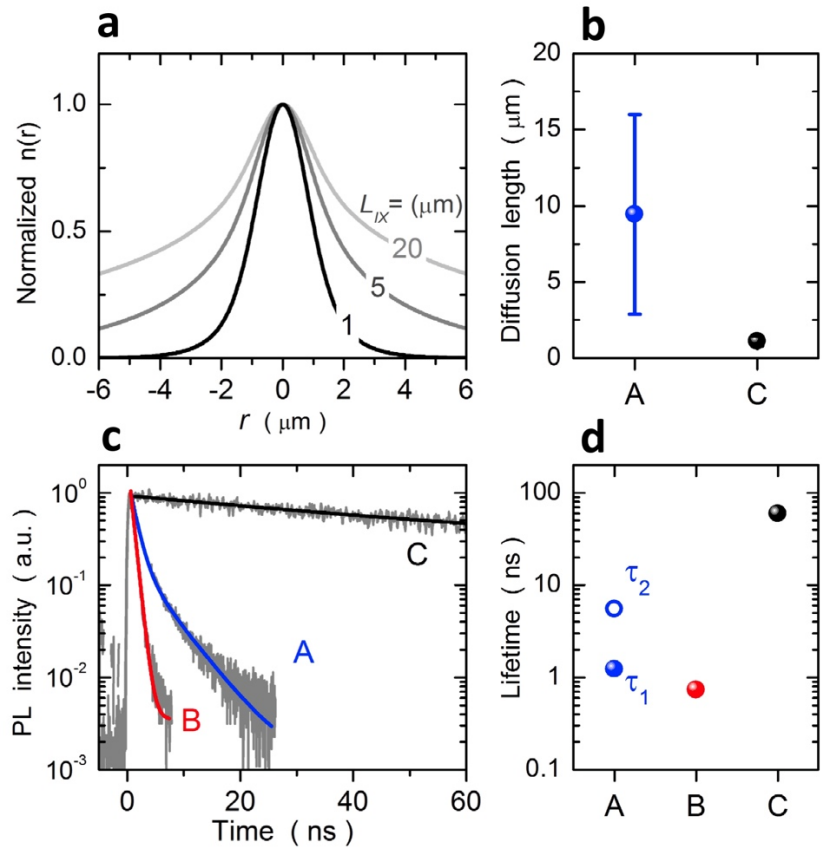

**Fig. 3. Determination of *IX*s diffusion length ($L_{IX}$) and measured *IX* lifetimes.** (a) Calculated spatial distribution of *IX* density in which n(r) is convoluted by the Gaussian laser profile with ~ 1 μm FWHM spot size. The measurements are compared to the simulated density distribution function to obtain the diffusion length for each sample. (b) *IX* diffusion length $L_{IX}$ from sample *A* and C. (c) TRPL spectra from sample *A*, *B*, and *C*, respectively. The TRPL spectrum was taken at the center of the *IX* resonances. The TRPL spectrum from sample *A* is fitted by a double-exponential decay function while sample *B* and *C* are fitted by a single-exponential function. The lifetimes extracted from fittings in (c) are summarized in (d) for all three samples. Note the error bars of lifetimes are smaller than the dots.

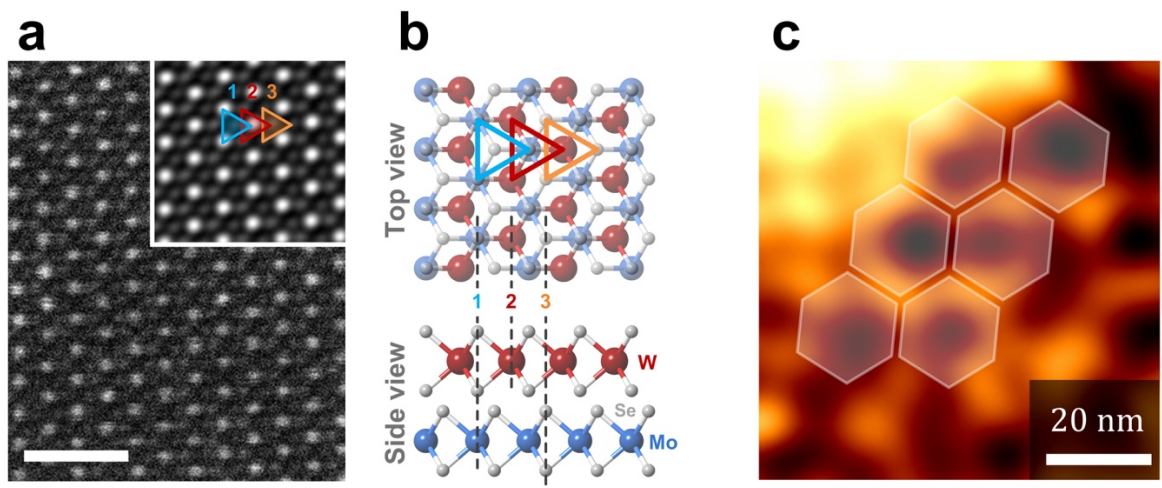

**Fig. 4. Crystal structures of CVD-grown and MET hBLs.** (a) A STEM image of a CVD-grown hBL similar to sample *A* shows a commensurate atomic registry, where the Bragg-filtered image is shown in the inset for comparison. The scale bar is 1 nm. (b) Schematic showing the ideal atomic registry with top (side) view in the upper (lower) panel. The STEM contrast originates from the difference of total atomic number in each atomic column. The different atomic columns are indicated by black dashed lines: column 1), 1 Mo atom and 2 Se atoms (blue triangle), column 2), 1 W atom (red triangle), and column 3), 2 Se atoms (orange triangle). (c) Filtered AFM phase image of an hBL prepared similarly to sample *B* and *C*, showing a hexagonal pattern consistent with the moiré superlattice. The bright spots at the upper left corner are from surface contaminations.